\documentclass[aps,preprint,amsmath]{revtex4}

\def\al{\alpha}
\def\be{\beta}
\def\ga{\gamma}

\def\ep{\epsilon}

\def\th{\theta}

\def\ka{\kappa}
\def\la{\lambda}

\def\rh{\rho}

\def\up{\upsilon}

\def\Ga{\Gamma}

\def\Si{\Sigma}

\def\cP{{\cal P}}
\def\cR{{\cal R}}

\def\mn{{\mu\nu}}

\def\fr#1#2{{{#1} \over {#2}}}
\def\frac#1#2{{\textstyle{{#1}\over {#2}}}}
\def\half{{\textstyle{1\over 2}}}

\def\tr#1{{\rm tr}#1}

\def\lsim{\mathrel{\rlap{\lower4pt\hbox{\hskip1pt$\sim$}}
    \raise1pt\hbox{$<$}}}
\def\gsim{\mathrel{\rlap{\lower4pt\hbox{\hskip1pt$\sim$}}
    \raise1pt\hbox{$>$}}}
\def\sqr#1#2{{\vcenter{\vbox{\hrule height.#2pt
         \hbox{\vrule width.#2pt height#1pt \kern#1pt
         \vrule width.#2pt}
         \hrule height.#2pt}}}}

\def\prt{\partial}
\def\pt#1{\phantom{#1}}

\def\etal{{\it et al.}}

\def\lrpartial{\raise 1pt\hbox{$\stackrel\leftrightarrow\partial$}}

\newcommand{\beq}{\begin{equation}}
\newcommand{\eeq}{\end{equation}}
\newcommand{\bea}{\begin{eqnarray}}
\newcommand{\eea}{\end{eqnarray}}
\newcommand{\rf}[1]{(\ref{#1})}
\newcommand{\nn}{\nonumber\\}

\def\rif{Riemann-Finsler}
\def\prif{pseudo-Riemann-Finsler}
\def\rie{Riemann}
\def\prie{pseudo-Riemann}
\def\fin{Finsler}
\def\pfin{pseudo-Finsler}
\def\ran{Randers}
\def\pran{pseudo-Randers}
\def\min{Minkowski}
\def\ber{Berwald}

\def\gram#1#2{{\rm gram}(#1,#2)}
\def\norm#1{\| #1 \|}

\def\tm#1{TM\backslash #1}

\def\m2{m^2}
\def\b2{b^2}

\def\u2{u^2}
\def\au{a\cdot u}
\def\bu{(b\cdot u)}

\def\up2{u_\perp^2}

\def\yy{y}
\def\y2{y^2}
\def\ay{a\cdot y}
\def\by{(b\cdot y)}
\def\ypa{y_\parallel}
\def\ype{y_\perp}
\def\yp2{y_\perp^2}

\def\ss{r}
\def\tt{s}
\def\ssi{\rh}
\def\face{{\it face}}
\def\kk{{B}}

\def\fa{{F_a}}
\def\fb{{F_b}}
\def\fab{{F_{ab}}}
\def\fsb{{(\fb)_{\Si \ka}}}

\def\vpa{v_\parallel}
\def\vpe{v_\perp}

\def\xj{{x_j}}
\def\xk{{x_k}}
\def\xl{{x_l}}

\def\yj{{\yy_j}}
\def\yk{{\yy_k}}
\def\yl{{\yy_l}}

\def\byp#1#2{b^{{#1}{#2}}}
\def\bdo#1#2{\tt_{{#1}{#2}}}

\def\tch#1#2#3{\widetilde{\ga}_{#1 #2 #3}}
\def\hch#1#2#3{\widehat{\ga}_{#1 #2 #3}}
\def\chr#1#2#3{\Ga_{#1 #2 #3}}
\def\chru#1#2#3{\Ga^{#1}_{\pt{#1} #2 #3}}

\def\tchu#1#2#3{\widetilde{\ga}^{#1}_{\pt{#1} #2 #3}}
\def\hchu#1#2#3{\widehat{\ga}^{#1}_{\pt{#1} #2 #3}}

\def\dt{\bullet}
\def\odt{\circ}

\def\cov{\widetilde D}

\begin{document}

\title{Riemann-Finsler geometry and Lorentz-violating kinematics}

\author{V.\ Alan Kosteleck\'y}

\affiliation{Physics Department, Indiana University,
Bloomington, IN 47405, U.S.A.} 

\date{IUHET 557, April 2011}

\begin{abstract}

Effective field theories with explicit Lorentz violation
are intimately linked to Riemann-Finsler geometry.
The quadratic single-fermion restriction of the 
Standard-Model Extension
provides a rich source of pseudo-Riemann-Finsler spacetimes 
and Riemann-Finsler spaces.
An example is presented that is 
constructed from a 1-form coefficient 
and has Finsler structure complementary to the Randers structure.

\end{abstract}

\maketitle

\section{Introduction}

The study of \rif\ geometry,
which has roots in Riemann's 1854 
{\it Habilitationsvortrag} 
and Finsler's 1918 dissertation 
\cite{br,pf},
is now an established mathematical field
with a variety of physical applications.
A well-known example intimately linked to physics 
is \ran\ geometry 
\cite{gr},
in which the \rie\ metric at each point is augmented
by a contribution from a 1-form.
For instance,
a \pran\ metric on (3+1)-dimensional spacetime
can be identified with the effective metric experienced 
by a relativistic charged massive particle
minimally coupled to a background electromagnetic 1-form potential.

The present work concerns 
the relationship between a large class of \rif\ geometries and
theories with explicit Lorentz violation.
Tiny Lorentz violation offers a promising prospect 
for experimental detection of new physics from the Planck scale
and could arise in an underlying unified theory such as strings
\cite{ksp}.
At attainable energies,
effective field theory provides a useful tool
for describing observable signals of Lorentz and CPT violation
\cite{kp,owg},
with explicit Lorentz violation characterized 
by background coefficients.
However,
explicit Lorentz violation 
is generically incompatible 
with the Bianchi identities of \prie\ geometry
\cite{akgrav}
and so presents an obstacle
to recovering the usual geometry of General Relativity.
This problem can be avoided via spontaneous Lorentz breaking as, 
for example,
in cardinal gravity
\cite{cardinal}.
An alternative might be to subsume
the usual \rie\ geometry 
into a more general geometrical structure.
Here, 
this is taken to be \rif\ geometry,
and one method is provided to connect it 
with Lorentz-violating effective field theories.
The notion of distance in \rif\ spaces and \prif\ spacetimes 
is controlled by additional quantities 
beyond the \rie\ metric
(for textbook treatments see, 
e.g., Refs.\ \cite{bcs,cs,haz,bf,ma}).
Intuitively,
the role of these quantities can be played by
the background coefficients for explicit Lorentz violation.

The comprehensive realistic effective field theory
with Lorentz violation
that incorporates both the Standard Model and General Relativity
is known as the Standard-Model Extension (SME)
\cite{dcak,akgrav}.
To relate the SME to \prif\ spacetimes and \rif\ spaces,
this work adopts as a starting point 
the single-fermion renormalizable restriction
of the SME in \min\ spacetime,
which is a comparatively simple quantum field theory
with explicit Lorentz violation.
In the presence of fermion self-interactions,
a connection between SME coefficients and \prif\ geometry
has been proposed by Bogoslovsky 
\cite{gyb}.
Here,
attention is focused on a free SME fermion,
which has a wave packet propagating
with a dispersion relation modified by Lorentz violation.
The dispersion relation is a quartic 
in the plane-wave 4-momentum $p_\mu$
and is exactly known
\cite{akrl}.
Some of its properties have been discussed by Lehnert
\cite{rl-fermion}
and by Altschul and Colladay
\cite{badc}.
It can be associated with an action 
for a relativistic point particle
\cite{aknr},
which encodes in a classical description
part of the key physical content of the free quantum field theory
while avoiding some of the complications associated with spin.

The basic observation underlying the present work
is that the SME-based Lorentz-violating classical lagrangian 
plays the role of a \pfin\ structure,
which leads to some interesting geometrical consequences.
Pseudo-\fin\ structures play a role 
in the context of modified particle dispersion relations 
\cite{gls}. 
They are also relevant 
for modified photon dispersion relations
\cite{lld},
for which the general photon dispersion relation 
arising from operators of arbitrary dimension is known
\cite{km-nonmin}.
A substantial recent literature 
links Lorentz violation with \prif\ geometries
in the contexts of spacetime, gravity, and field theory
\cite{gsa,sv,zcxl,hg,ls,kss,nm,jsmv,nv,pw}.
Note,
however,
that at present no compelling experimental evidence
exists for Lorentz violation in nature 
\cite{tables},
although SME-based models provide
simple explanations for certain unconfirmed experimental results
including anomalous neutrino oscillations
\cite{jdak}
and anomalous meson oscillations
\cite{kvk}.

Interpreting the classical SME lagrangian as a \pfin\ structure 
implies that the Lorentz-violating trajectories
of the relativistic particles
are governed by \prif\ metrics
and hence define \prif\ geometries
in (3+1) dimensions. 
Corresponding \rif\ geometries
can be generated by Wick rotation 
or restriction to spatial submanifolds.
This construction can be extended 
to arbitrary dimensional curved spaces
in a straightforward way.
Following a general discussion of these ideas,
this work presents some basic results for a particular \rif\ space
that is constructed using a \rie\ metric and a 1-form 
but differs from \ran\ space for dimensions $n\ge 3$.
Its \fin\ structure is complementary to the \ran\ structure 
in a certain sense described below.

The conventions used here are as follows.
Coordinates for ($n$+1)-dimensional \prif\ spacetime
are denoted $x^\mu$, $\mu = 0,1,\ldots, n$.
The velocities $u^\mu$ along a curve
with path parameter $\la$ are $u^\mu = dx^\mu/d\la$,
and a \pfin\ structure is denoted as $L=L(x,u)$.
The \min\ metric 
in ($n$+1) dimensions is defined 
to have positive signature for $n>2$.
Index raisings or lowerings
and contractions are performed 
with the \prie\ metric $\ss_{\mu\nu}$
and its inverse $\ss^{\mu\nu}$;
for example,
$u_\mu = \ss_{\mu\nu} u^\nu$ and
$\u2 = u^\mu \ss_{\mn} u^\nu$.
The \prif\ metric is $g_{\mu\nu}$ with inverse $g^{\mu\nu}$.
To match conventions in mathematics 
(see, e.g., Ref.\ \cite{bcs}),
coordinates for $n$-dimensional \rif\ space
are denoted $x^j$, $j = 1,\ldots, n$,
the velocities are $\yy^j=dx^j/d\la$,
and a \fin\ structure is denoted as $F=F(x,y)$.
The \rie\ metric is $\ss_{jk}$
with inverse $\ss^{jk}$,
while the \rif\ metric is $g_{jk}$ with inverse $g^{jk}$.
Index raisings or lowerings
and contractions are performed 
with the \rie\ metric;
for example,
$\yy_j = \ss_{jk} \yy^k$,
$\y2 = \yy^j \ss_{jk} \yy^k$.
The norm $\norm \yy$ of $\yy^j$ is $\norm \yy = \sqrt{\y2}$.
Partial derivatives with respect to $y^j$
are denoted by subscripts;
for example, 
$\prt F/\prt y^j = F_{y^j}$.

\section{SME-based \fin\ structures}

The Lagrange density for 
the renormalizable single-fermion restriction 
of the minimal SME in (3+1)-dimensional \min\ spacetime
includes all operators quadratic in the fermion field
having mass dimensions three and four
\cite{dcak}.
Each Lorentz-violating operator is contracted 
with a controlling coefficient,
so the physics is coordinate independent.
The coefficients of mass dimension one 
are conventionally denoted as
$a_\mu$, $b_\mu$, $H_\mn$,
while the dimensionless ones are
$c_\mn$, $d_\mn$, $e_\mu$, $f_\mu$, and $g_{\la\mu\nu}$.
A constructive procedure has recently been given for generating 
the classical relativistic point-particle lagrangian $L$
from which the SME plane-wave dispersion relation can be derived
\cite{aknr}.
The complete action is involved,
but various special cases are tractable 
and some limits of $L$ have been explicitly obtained.
One example of relevance in what follows
is the limiting situation of vanishing coefficients
$c_\mn$, $d_\mn$, $e_\mu$, $f_\mu$, $g_{\la\mu\nu}$,
and $H_\mn$,
for which the particle lagrangian is 
\bea
L_{ab} =
- m \sqrt{-\u2} - \au \mp \sqrt{\bu^2 - \b2 \u2} .
\hskip -20pt
\label{ab}
\eea
The two possible signs for the last term reflect
the presence of two particle spin projections
in the quantum field theory.

This work extends the full $L$ and its limits to include
minimal coupling to a background gravitational field 
given by a \prie\ metric $\ss_{\mu\nu}$
in ($n$+1) dimensions
and to allow position dependence of all coefficients.
For example,
this extension affects
the contractions in Eq.\ \rf{ab}
and permits $L_{ab}$ to describe the motion 
of a relativistic particle 
on an ($n$+1)-dimensional curved spacetime manifold
in the presence of varying background coefficients 
$a_\mu(x)$ and $b_\mu(x)$.
The extended $L$ could be obtained via a suitable
Foldy-Wouthuysen transformation
\cite{fw}
of the gravitationally coupled Dirac equation.
Position dependence of SME coefficients
appears naturally in the gravity context
\cite{akgrav,qbakpn,abk,lvmm},
and Seifert has shown this can result 
from topologically nontrivial field configurations
\cite{mds}.
Note that comparatively large $a_\mu$ coefficients 
could have escaped experimental detection to date
\cite{akjt}.
In what follows the fermion mass $m$ is set to unity, $m=1$,
for simplicity.

The classical relativistic lagrangian can be viewed 
as a function $L=L(x,u)$
on the tangent bundle $TM$ 
of the background spacetime manifold $M$.
The Lorentz violation is assumed sufficiently small
that nonzero values of $L$ have only one sign,
fixed by the mass term.
The function $L$ is smoothly differentiable everywhere 
except along a subset $S=S_0 + S_1$ of $TM$ 
that includes the usual slit $S_0$ with $u^\mu = 0$
and possibly also an extension $S_1$.
The requirement of curve-reparametrization invariance
imposes positive homogeneity of $L$ of degree one in $u^\mu$:
$L(x,\ka u) = \ka L(x, u)$ for $\ka > 0$.
The Lorentz violation is also assumed sufficiently small 
that the nonsingular \prie\ metric dominates the background fields,
so the effective metric 
$g_{\mn} := \prt^2 (L^2/2)/\prt u^\mu\prt u^\nu$
felt by the relativistic particle is nonsingular.
Inspection reveals that the above results 
are the defining properties
of a local \prif\ spacetime
with \pfin\ structure $L$ defined on $\tm S$
(for a textbook discussion see, e.g., Ref.\ \cite{bf}).
This \prif\ spacetime therefore underlies the motion 
of a relativistic classical particle experiencing
general Lorentz violation.

The explicit forms of $L$ and of many of its limits
are involved and unknown in detail,
so the corresponding \pfin\ geometries 
may be challenging to explore.
However,
a variety of special \prif\ spacetimes can be obtained
by taking limits in which certain coefficients vanish.
One simple example is 
the structure $L_a := L_{ab}|_{b\to 0}$
obtained as the limit $b_\mu \to 0$ of Eq.\ \rf{ab},
which takes the familiar \pran\ form.
The `\face' limit $L_{acef}$ of $L$
with coefficients $a_\mu$, $c_\mn$, $e_\mu$, and $f_\mu$
also yields a \pran\ structure.

An interesting class of comparatively simple limits of $L$ 
consists of `bipartite' \pfin\ structures taking the generic form 
\bea
L_{\tt} =
- \sqrt{-\u2} 
\mp \sqrt{- u^\mu \tt_{\mn} u^\nu} ,
\label{mp}
\eea
where the symmetric quantity $\tt_{\mu\nu}$ satisfies
$u^\mu \tt_{\mn} u^\nu\leq 0$.
Several of the more tractable limits of $L$ 
fall into this class.
For example,
the $b$ structure 
$L_b := L_{ab}|_{a\to 0}$ is bipartite, 
with $\tt_{\mn} = \b2 \ss_{\mn} - b_\mu b_\nu $.
The example with $H_{\mn}$ 
given in Eq.\ (15) of Ref.\ \cite{aknr}
is also a bipartite structure $L_H$.
The choice $\tt_{\mu\nu} = - a_\mu a_\nu$
yields the two structures $L_{|a|} = - \sqrt{-\u2} \mp | \au |$
jointly spanning $L_a$,
so $L_{ab}$ has a tripartite form in this sense.
It is likely that other bipartite limits of $L$
remain to be discovered. 
Note that the quantity $\tt_{\mu\nu}$ 
is reminiscent of a secondary metric but may lack an inverse.
For instance,
$\tt_{\mn}$ for the $b$ structure $L_b$ is noninvertible
because it has a zero eigenvalue for the eigenvector $b^\mu$.
As a result,
the subset $S$ for this example
consists of the extended slit $u^\mu = \ka b^\mu$ for real $\ka$.
In the more general case,
$S_1$ includes all $u^\mu$ that are nonzero eigenvectors of 
$\tt_{\mu\nu}$ with zero eigenvalues.

When $S=S_0$,
which holds for the \face\ structure
and other \pran\ limits of $L$,
global \pfin\ spacetimes arise.
However,
when $S_1$ is nonempty,
the geometry is only local.
An interesting open question is whether
it is possible to resolve the geometry at $S_1$ 
to yield global \pfin\ spacetimes.
The fourth-order polynomial dispersion relation 
for the wave-packet 4-momentum $p_\mu$
can be viewed as an algebraic variety $\cR(p_\mu)$.
The structure $L$ is constructed using $\cR$,
the requirement of homogeneity,
and the intrinsic derivatives of $L$ 
defining the 4-velocity $u^\mu$
\cite{aknr}.
This construction generates five equations
that combine to yield a polynomial $\cP(L)$,
which has physical roots yielding  
the local \pfin\ structures $L$ of interest
and spurious roots corresponding to the set $S_1$.
The latter arise from the singularities of $\cR$,
which according to the implicit function theorem
are determined by the $p_\mu$ derivatives of $\cR$.
Resolving the geometry at $S_1$ therefore corresponds
to resolving the singularities of the variety $\cR$. 

At the level of quantum field theory,
the singularities of $\cR$ reflect degeneracy
of the wave-packet energies,
which can be resolved using spin.
For example,
in the Lorentz-invariant case
the two spin projections for the particle modes 
are degenerate for all momenta 
because the variety $\cR = (p^2 + 1)^2$ is singular everywhere,
but considering only one spin projection at a time
yields the nonsingular variety $\cR = p^2 + 1$ instead.
This spin-based resolution also underlies
the global nature of the \pran\ \face\ geometry.
When any of $b_\mu$, $d_\mn$, $g_{\la\mu\nu}$, or $H_\mn$
is nonzero,
$\cR$ is generically a nontrivial quartic.
Singularities occur for a subset of momenta 
at which the two spin projections are degenerate in energy,
and these generate the set $S_1$ in $TM$.
It is therefore plausible that the \pfin\ geometry at $S_1$
for the general structure $L$
could be resolved by the introduction of a spin variable.
Note that a resolution for the corresponding variety $\cR$ 
is guaranteed by,
for example,
the Hironaka theorem \cite{hh}.
The geometry at $S_1$ might therefore alternatively be resolved 
using a standard technique 
for singularities of algebraic varieties
such as blowing up.
The above comments suggest 
the existence of a global \pfin\ geometry associated 
with the general structure $L$
is a reasonable conjecture,
but its proof remains open. 

An interesting class of SME-based \fin\ structures 
can be obtained from the \pfin\ ones
by restriction to the spatial submanifold or by Wick rotation.
The full \fin\ structure $F(x,y)$ obtained 
in this way retains much of the complexity of $L(x,u)$,
but some limits are amenable to explicit investigation.
One comparatively simple example
arises by converting to $n$ euclidean dimensions
the \pfin\ $ab$ structure $L_{ab}$ given in Eq.\ \rf{ab},
yielding the \fin\ $ab$ structure
\bea
F_{ab} =
\sqrt{\y2} + \ay \pm \sqrt{\b2 \y2 - \by^2} .
\eea
Note that $\fa := F_{ab}|_{b\to 0}$ 
generates the usual \ran\ geometry.
The \fin\ \face\ structure $F_{acef}$
corresponding to $L_{acef}$ 
also generates a \ran\ space. 

Applied to the bipartite \pfin\ structure \rf{mp},
the above procedure generates
a bipartite \fin\ structure $F_{\tt}$
given by
\bea
F_{\tt} =
\sqrt{\y2} \pm \sqrt{y^j \tt_{jk} y^k} ,
\label{finbip}
\eea
where the contractions now involve 
a positive-definite \rie\ metric $\ss_{jk}$.
For the lower sign choice,
the nonnegativity of $F_{\tt}$
implies $\tt_{jk}$ must be bounded
and yields the constraint
$\det{(1 - \ss^{-1} \tt)} > 0$.
This corresponds to the assumption that the Lorentz violation
is perturbative.
Among bipartite examples are the \fin\ $b$ structure
$\fb := F_{ab}|_{a\to 0}$ 
and the $H$ structure $F_H$ obtained
by restricting the \pfin\ structure $L_H$ 
for the coefficient $H_\mn$ to spatial components.
With the positive sign in Eq.\ \rf{finbip}
and invertible $\tt_{jk}$,
$F_{\tt}$ reduces to the two-metric $y$-global \fin\ structure 
mentioned by Antonelli, Ingarden, and Matsumoto 
in the context of photon birefringence in uniaxial crystals
(see Eq.\ (4.2.29) of Ref.\ \cite{aim}).
In the present fermion context,
however,
$\det{\tt}\geq 0$ can vanish
and hence $\tt_{jk}$ may have no inverse,
implying a nonempty slit extension $S_1$.
Indeed,
the \fin\ structures $F$ and its limits
associated with $L$ 
are typically $y$-local,
although global \pran\ structures yield global \ran\ structures.
It is plausible that the putative $y$-global completions 
of \pfin\ structures discussed above  
would also yield $y$-global \rif\ submanifolds.

\section{The $b$ structure}

As an explicit example of an SME-based \rif\ geometry,
consider the $b$ structure $\fb := F_{ab}|_{a\to 0}$ 
obtained as a limit of Eq.\ \rf{finbip}.
In fact this specifies two \fin\ structures,
one for each choice of the $\pm$ sign,
originating in the two spin degrees of freedom in the SME.
For notational simplicity,
it is convenient to write the $ab$ structure $\fab (x,y)$ as 
\beq
\fab = \ssi + \al + \be,
\label{bfs}
\eeq
where
\beq
\ssi := \sqrt{\y2},
\quad
\al := \ay,
\quad
\be := \pm \sqrt{\b2 \y2 - \by^2}.
\eeq
For a \rie\ space with metric $\ss_{jk}$,
the \fin\ structure is $F_{\ss} = \ssi$,
for \ran\ $a$ space it is $\fa = \ssi + \al$,
and for $b$ space it is $\fb = \ssi + \be$.
The dependence on $x^j$ arises through $\ss_{jk}(x)$, $a_j(x)$,
and $b_j(x)$.
Constancy of the metric and coefficients
would imply that the canonical momentum is conserved
and that the \rif\ space is locally \min,
which parallels the treatment of Ref.\ \cite{aknr}.
The notational pairings $(\ss,\ssi)$, $(a,\al)$, $(b,\be)$ here
match the standard literature on Lorentz violation;
the conventional mathematics notation
is recovered by the replacements 
$(\ss,\ssi)\to (a,\al)$, $(a,\al)\to (b,\be)$,
$(b,\be)\to (\star,\star)$. 

A first observation is that the $b$ structure $\fb$
offers a kind of complement to the \ran\ $a$ structure $\fa$.
Given a nonzero 1-form $a_j$,
$\fa$ can be constructed
by adding to $F_{\ss}$ the parallel projection  
of the velocity $\yy^j$ along $a_j$,
\beq
\fa = \ssi + \al = \sqrt{\y2} + \norm{a}~\ypa ,
\eeq
where $\ypa = \ay/\norm{a}$ is
the $a$-normalized parallel projection.
By splitting the \ran\ structure into two pieces,
the last term can be written
$\al= \pm \norm a \sqrt{\ypa^2}$.
However,
given another nonzero 1-form $b_j$,
a complementary structure can be obtained
by combining $F_{\ss}$ 
with the perpendicular projection
of the velocity $\yy^j$ along $b_j$ instead.
This gives the $b$ structure $\fb$,
\beq
\fb = \ssi + \be = \sqrt{\y2} \pm \norm b \sqrt{\yp2},
\label{b}
\eeq
where $\ype^j = \yy^j - \by b^j/\b2$ is
the $b$-normalized perpendicular projection.
One natural formulation of this perpendicular projection 
uses the Gram determinant or gramian. 
Given two vectors $b^j$, $\yy^j$ and the \rie\ metric $\ss_{jk}$,
the gramian $\gram b u$ is given by 
$\gram b u = \b2 \y2 - \by^2$,
so the $b$ structure can be written 
\beq
\fb = \sqrt{\y2} \pm \sqrt{\gram b \yy}.
\label{gram}
\eeq
In euclidean space,
the gramian of two vectors represents the square of the area 
of the parallelogram formed by the vectors.
The Cauchy-Schwarz inequality implies
the gramian is a nonnegative quantity, 
$\gram b \yy \geq 0$,
confirming that the square-root term in Eq.\ \rf{gram} is real,
as required.

For low dimensions,
the $b$ structure $\fb$ generates known geometries.
When $n=1$ the gramian vanishes,
$\gram b \yy = 0$,
so the \rif\ space reduces to a \rie\ curve.
When $n=2$,
the parallel and perpendicular projections
$\ypa^j$ and $\ype^j$ span vector spaces of the same dimension.
This enables the introduction of a vector $v^j$
via the identification $\ype^j\to\vpa^j$, $\ypa^j\to\vpe^j$,
which maps 
$\ssi(\yy) \to \ssi (v)$ and
$\be = \pm \norm b \sqrt{\yp2} \to \pm \norm b \sqrt{\vpa^2}$.
The $b$ structure with its two signs therefore maps 
to the two pieces of a \ran\ structure.
An equivalent way to see this result
is to identify the corresponding \ran\ 1-form $a_j$
with the dual of $b_j$,
$a_j = \ep_{jk}b^k$,
which is perpendicular to $b_j$.
The contribution to $\fb$
from the perpendicular projection to $b_j$
is equivalent to a contribution to $\fb$
from the parallel projection to the dual 
$\ep_{jk}b^k$,
so for $n=2$ the $b$ structure generates a \ran\ geometry.
However,
the duality equivalence 
is unavailable in higher dimensions,
so for $n\geq 3$ 
the $b$ space is expected to be 
neither a \rie\ nor a \ran\ geometry.
This result is proved in the next section 
by direct construction of the Matsumoto torsion.

To be a \fin\ structure,
$\fb$ must satisfy certain basic criteria
\cite{bcs}.
One is nonnegativity on $TM$.
For the positive sign in Eq.\ \rf{b}, 
$\fb$ is always nonnegative because $\ssi \geq 0$ 
and $\be \geq 0$.
For the negative sign,
$\fb \geq 0$ iff $\norm b < 1$.
This can be checked as follows.
If $b_j$ is zero then $\fb = \ssi$,
which is nonnegative.
If $\yy^j$ is zero then $\fb = 0$,
which is also nonnegative. 
If both $b_j$ and $\yy^j$ are nonzero,
define the nonzero real angle 
$\cos \th = \by/(\norm b ~ \norm \yy)$.
Then $\fb = \norm \yy (1 \pm \norm b ~ | \sin\th | )$.
So if $\fb_- >0$ then $\norm b <1$ 
because $0\leq | \sin\th | \leq 1$.
Also,
if $\norm b < 1$ then $\fb_- >0$ for the same reason.
The nonnegativity of $\fb$ is therefore assured 
for both signs in $\fb$ when $\norm b <1$.
This condition is assumed in what follows.

Another criterion for a \fin\ structure is $C^\infty$ regularity.
Since the \rie\ metric is positive definite,
the component $\ssi$ of $\fb$ is $C^\infty$ 
on the usual slit bundle $\tm{S_0}$ for which $\yy^j \neq 0$.
In contrast,
the component $\be$ vanishes on the slit extension $S_1$
for which $\ype^j = 0$ and $\yy^j \neq 0$,
so on $\tm{S_0}$ only $C^0$ continuity of $\fb$ is assured
in the general case.
However,
$\be$ is positive definite 
outside the set $S=S_0+S_1$ for which $\gram b \yy = 0$.
This implies that $\fb$ is $C^\infty$ on $\tm S$.
Where necessary, 
the restriction of $\fb$ 
to $\tm S$ is assumed in what follows.
As discussed in the previous section,
when $S_1$ is nonempty this restriction implies 
the geometry associated with $\fb$ is typically singular on $S$
and hence is $y$ local.
Exceptions are the case $n=1$,
which generates a \rie\ curve and is $y$ global,
and the case $n=2$,
which can be mapped to a $y$-global \ran\ geometry
as described above.
The singularities at $\gram b \yy = 0$, $\yy^j \neq 0$
originate in those at $\gram b u = 0$, $u^j \neq 0$
arising from the \pfin\ structure $L_b$.
In turn,
these are associated with singularities
of the algebraic variety $\cR$
mentioned in the previous section.
Some calculation shows the latter appear at
$\gram b p = 0$, $p^\mu \neq 0$,
where the dispersion relation has solutions
with degenerate energies for spin projections
satisfying $p_\mu = \pm \sqrt{(1 + \m2/\b2)} ~b_\mu$
for timelike $b_\mu$.
Colladay, McDonald, and Mullins
have exhibited the dispersion relation 
as intersecting pairs of deformed spheres 
\cite{cmm}.
In projection, 
the degenerate energies appear as cusps on the energy-momentum plot
\cite{akrl}.
Resolving these singularities 
and generating the corresponding $y$-global \rif\ geometries
for $\fb$ is an interesting open problem.

The two remaining criteria for $\fb$
to be a \fin\ structure 
are positive homogeneity of degree one in $\yy^j$,
$\fb (x,\ka \yy) = \ka \fb (x, \yy)$ for $\ka > 0$,
and positive definiteness of the symmetric \fin\ metric
$g_{jk} := (\fb^2/2)_{\yj\yk}$
associated with $\fb$.
The former holds by inspection,
but to demonstrate the latter
some explicit results are useful.

A short calculation shows $g_{jk}$ 
can be expressed compactly as
\bea
g_{jk} &=&
\fr \fb \ssi \fr \kk \be \ss_{jk}
- \ssi\be \ka_j \ka_k
- \fr \fb \be b_j b_k,
\quad
\label{metric}
\eea
where $\kk := \be + \b2 \ssi$ 
and where $\ka_j$ represents the convenient combination 
\bea
\ka_j &:=& \fr{\ssi_\yj}{\ssi} - \fr{\be_\yj}{\be}
\eea
involving the $\yy^j$ derivatives of $\ssi$ and $\be$.
The latter are 
$\ssi_\yj = {\yy_j}/ \ssi$
and 
$\be_\yj = \tt_{jk} \yy^k/ \be$,
where $\tt_{jk} = \b2 \ss_{jk} - b_j b_k$
for the $b$ structure.
One way to investigate positive definiteness of $g_{jk}$ 
is via the determinant $\det g$.
For $n=1$ the determinant is $\det g = \det \ss$,
matching expectations for a \rie\ curve.
For arbitrary $n\geq 2$,
some calculation gives the pleasantly simple formula
\bea
\det g &=&
\left(\fr \kk \be \right)^{n-2}
\left( \fr {\fb} {\ssi} \right)^{n+1} \det \ss.
\label{det}
\eea
For $n=2$, 
the first factor reduces to the identity
and the remaining factors match the well-known
determinant of the \ran\ metric,
as might be expected
from the $n=2$ mapping between the $a$ and $b$ structures.
Also,
in the limit $\norm b \to 0$
the formula produces $\det g = \det \ss$,
as required.

Given the result \rf{det},
a standard argument
\cite{bcs}
verifies positive definiteness of $g_{jk}$.
Introducing $F_{\ep b} = \ssi + \ep \be$,
it follows from \rf{det}
that $\det g_\ep$ is positive 
and so ${g_\ep}_{jk}$ has no vanishing eigenvalues.
At $\ep=0$ the eigenvalues of ${g_\ep}_{jk}$
are those of $\ss_{jk}$ and hence are all positive,
while as $\ep$ increases to 1 no eigenvalue can change sign
because none vanishes. 
This ensures positive definiteness 
and also invertibility of $g_{jk}$.

\section{Some properties of $b$ space}

For any $n>1$,
the \fin\ $b$ space with metric \rf{metric} 
cannot be a \rie\ geometry.
One way to see this is to construct the 
the Cartan torsion 
$C_{jkl} := (g_{jk})_{\yy^l}/2$,
which measures the non-euclidean nature
of a \fin\ structure viewed as a \min\ norm 
on any tangent space $TM_x$.
For $b$ space,
the Cartan torsion takes the simple form 
\bea
C_{jkl} &=&
- \half\ssi\be \sum_{(jkl)} \ka_j\ka_{kl},
\eea
where the sum is over cyclic permutations of $j$, $k$, $l$.
Here,
$\ka_{jk}$ is the combination
\beq
\ka_{jk} := \fr {\ssi_{\yj\yk}} {\ssi} - \fr{\be_{\yj\yk}} {\be}
\eeq
of the second $\yy^j$ derivatives of $\ssi$ and $\be$,
which are
$\ssi_{\yj\yk} = (\ss_{jk} - \ssi_\yj \ssi_\yk)/\ssi$
and 
$\be_{\yj\yk} = (\tt_{jk} - \be_\yj \be_\yk)/\be$.
Note that $\be_{\yj\yk}$ vanishes for $n=2$.
Since $C_{jkl}$ is nonzero,
Diecke's theorem
\cite{diecke}
implies that $\fb$ is non-euclidean as a \min\ norm,
so $b$ space cannot be a \rie\ geometry.
The mean Cartan torsion 
$I_j := (\ln( \det g))_{\yy^j}/2$
is found to be
\bea
I_j &=&
-\half 
\left[ (n+1) \fr{\be}{\fb} 
- (n-2) \fr{\b2\ssi}{\kk} \right]\ka_j, 
\eea
which is also nonvanishing for $n>1$.

For any $n>2$,
the $b$ space also differs from \ran\ space. 
This can be seen by calculating 
the Matsumoto torsion $M_{jkl}$,
which separates \ran\ and non-\ran\ metrics when $n>2$.
This torsion is defined as 
$M_{jkl} := C_{jkl} - \fr 1 {(n+1)} \sum_{(jkl)} I_j h_{kl}$,
where the angular metric $h_{jk}$ is 
$h_{jk} := g_{jk}- F_\yj F_\yk$.
For $\fb$,
the Matsumoto torsion can be written as 
\bea
M_{jkl} &=&
-\half \fb \sum_{(jkl)} \ka_j
\left[
 \fr{(n-2)}{(n+1)} \fr {\b2\ssi} \kk
( \ssi_{kl} + \be_{kl}) - \be_{kl}
\right].
\nn
\eea
Since this is nonzero for $n>2$,
the Matsumoto-H\=oj\=o theorem
\cite{mh}
shows that the $b$ structure $\fb$ 
cannot correspond to a \ran\ structure for $n > 2$,
despite being constructed from a 1-form $b_j$
and despite its comparative simplicity and calculability.

One way to explore features of a \rif\ space 
is to study its geodesics
(for a textbook treatment see, e.g., Ref.\ \cite{zs-sprays}).
The \fin\ geodesics for $b$ space are solutions of the equation
\bea
\fb \fr d {d\la} \left( \fr 1 \fb \fr {dx^j} {d \la} \right) 
+ G^j =0,
\label{fingeo}
\eea
where the spray coefficients $G^j := g^{jm} \chr mkl y^k y^l$
are defined in terms of the Christoffel symbol $\chr jkl$ 
for the \rif\ metric $g_{jk}$, 
\bea
\chr j k l &:=& 
\half (
\prt_\xk g_{jl} + \prt_\xl g_{jk} 
- \prt_\xj g_{kl}).
\eea
The geodesics solving Eq.\ \rf{fingeo}
are valid for any choice of diffeomorphism gauge or,
equivalently,
for any choice of geodesic speed.

The spray coefficients $G^j$ for $b$ space 
can be calculated explicitly
by first deriving $G_j := \chr jkl y^k y^l$
and then contracting with the inverse \rif\ metric to get
$G^j := g^{jk} G_k$.
Some calculation reveals the compact result 
\bea
G_j = \ssi \fb \tch j \dt \dt 
+ \ssi^2 (\prt_\dt \be - \be \tch \dt\dt\dt ) \ka_j
+ \fr {\ssi^2 \fb}{\be} \hch j \dt\dt.
\eea
Here,
a lower index $m$ contracted with $\ss^{mk}\ssi_\yk$ 
is denoted by a bullet $\dt$,
with contractions understood to be external to any derivatives.
Also,
the Christoffel symbol $\tch j k l$ 
for the \rie\ metric $\ss_{jk}$ takes the usual form
\bea
\tch j k l &:=& 
\half (
\prt_\xk \ss_{jl} + \prt_\xl \ss_{jk} 
- \prt_\xj \ss_{kl}),
\eea
while the symbol $\hch j k l$ 
is defined analogously as
\bea
\hch j k l &:=& 
\half (
\prt_\xk \bdo jl + \prt_\xl \bdo jk
- \prt_\xj \bdo kl)
\eea
using the form of $\bdo jk$ for $\fb$. 

To proceed,
the inverse \rif\ metric is required.
This can be determined to be
\bea
g^{jk} &=&
\fr \ssi \fb 
\left(
\ss^{jk}
+ \fr {\by^2 \ssi}{\kk \be^2} \la^j\la^k
- \fr{\ssi}{\kk } \byp jk
\right),
\eea
where
\bea
\la_j &:=& \fr{\by}{\fb} \ssi_\yj - b_j.
\eea
Contracting with $G_j$ gives the spray coefficients $G^j$ as
\bea
G^j &=& 
\ssi^2 \tchu j\dt\dt
+ \fr{\ssi^3}{\kk \be^3}
\big[
\be^3 \hchu j\dt\dt 
+ \ssi^2\be \hch \odt\dt\dt b^j
\nn
&&
\hskip 50pt
-\ssi \yy_\odt (\hch \odt\dt\dt + \be \hch \dt\dt\dt)\la^j)
\big],
\label{spray}
\eea
where a lower index $m$ contracted with $b^m$ 
is denoted by an open circle $\odt$.
This result implies that the geodesic equation on $b$ space
can be viewed as the usual \rie\ geodesic equation
corrected by terms involving
the symbol $\hch j k l$.

The expression \rf{spray} for the spray coefficients
leads to some insights about $b$ space.
Suppose the 1-form $b_j$ is parallel
with respect to the \rie\ metric $\ss_{jk}$,
$\cov_j b_k = 0$.
Then,
the \fin\ geodesics reduce to the standard \rie\ geodesics
for the metric $r_{jk}$. 
This can be demonstrated via the explicit formula 
\bea
\ss_{jk} \hchu k\dt\dt &=&
2 \ssi_\yj \cov_\dt b_\odt
- \cov_j b_\odt
\nn &&
- b_j \cov_\dt b_\dt
- b_\dt \cov_\dt b_j 
+ b_\dt \cov_j b_\dt,
\eea
where $\cov_j$ is the \rie\ covariant derivative
and contractions are understood to be external to derivatives,
as before.
It follows that if $\cov_j b_k = 0$
then $\hchu j\dt\dt = 0$
and so also $\hch \dt\dt\dt = \hch \odt \dt\dt = 0$.
The \fin\ spray coefficients \rf{spray} therefore become
$G^j = \ssi^2 \tchu j\dt\dt = \tchu jkl \yy^k \yy^l$,
which are the usual \rie\ spray coefficients
for the metric $\ss_{jk}$. 
For constant \fin\ speed
or, 
equivalently,
the gauge choice $\fb=1$ fixing the curve parameter $\la$
to a definite time $\la=t$,
the geodesics then become solutions of
the usual \rie\ geodesic equation
$\ddot x^j + \tchu jkl \dot x^k \dot x^l = 0$.

Remarkably,
this result shows an $\ss$-parallel $b_j$ coefficient 
has no effect on the motion.
Intuitively,
local conditions along the geodesics appear uniform,
so local geodesic observations cannot unambiguously detect 
nonzero $b_j$. 
This suggests 
a suitable transformation or coordinate redefinition
could be found to remove a parallel $b_j$ from $\fb$,
in analogy to the removal of certain unphysical coefficients 
in suitable limits of the SME 
\cite{akgrav,dcak,aknr,km-nonmin,akjt,redef,rl-redef}.
For example,
at least one component of the \ran\ coefficient $a_\mu$
can be removed by a phase redefinition of the fermion
\cite{akgrav}.
At the relativistic quantum level,
the $b_\mu$ coefficients cannot generically be removed 
due to the entanglement of the spin components,
which is absent at the classical level away from the set $S$.
However,
for constant $b_\mu$ in \min\ spacetime,
a chiral phase transformation can eliminate $b_\mu$
in the massless limit
\cite{dcak},
and Lehnert has exhibited a nonlocal field redefinition
that simultaneously removes $b_\mu$ from both spin components
\cite{rl-redef}. 

The expression \rf{spray} for the spray coefficients
permits in principle the direct derivation
of various geometric quantities
characterizing $b$ space,
including the nonlinear connection
$N^j_{\pt{j}k} := (G^j)_\yj/2$,
the \ber\ connection
$^B{}\chru j k l := (G^j)_{\yj\yk}/2$, 
and the \ber\ h-v curvature
$^B{}P_k{}^j{}_{lm} := - \fb (G^j)_{\yj\yk\yl}/2$.
The Cartan, Chern (Rund), and Hashiguchi connections
and the various associated curvatures and torsions 
can also in principle be obtained.
However,
the explicit formulae appear lengthy
and are omitted here.

One result of interest pertaining to \ber\ curvature
is that any $b$ space having $b_j$ parallel 
with respect to $\ss_{jk}$ is a \ber\ space.
Since $\cov_j b_k = 0$ implies $\hchu j\dt\dt = 0$
and hence $G^j = \tchu jkl \yy^k \yy^l$,
and since $\tchu jkl$ is independent of $\yy^j$, 
three $\yy$ derivatives of $G^j$ vanish.
The \ber\ h-v curvature is therefore zero,
and so any $\ss$-parallel $b$ space is a \ber\ space.
The converse statement that any \ber\ $b$ space 
is necessarily an $\ss$-parallel space 
appears plausible but is left open here.

Note that the analogous results for \ran\ space,
which in present terminology state that any $a$ space 
is a \ber\ space iff it is an $\ss$-parallel space,
are well established
\cite{mm,hi,ssay,sk}.
It is natural to conjecture that
any SME-based \rif\ space is a \ber\ space
iff it has $\ss$-parallel coefficients 
for Lorentz violation.
This attractive conjecture is amenable to direct 
investigation in various special cases,
while a general proof is likely to offer valuable insights.

Another open challenge is to identify physical interpretations
of SME-based \prif\ and \rif\ structures,
including the $b$ structure.
Examples for the $a$ structure are well known.
As mentioned in the introduction,
the dynamics of a relativistic charged particle 
moving in an electromagnetic potential
is governed by a \pran\ $a$ structure $L_a$,
while the \ran\ $a$ structure $\fa$ has applications
in several physical contexts
including 
Zermelo navigation,
optical metrics,
and magnetic flow
(see, e.g., Refs.\ \cite{zs,brs,dbcr,ghww,ts}).
Physical applications of the $b$ structure 
would also be interesting  
from both physical and mathematical perspectives.
By construction,
the SME-based \prif\ $b$ structure $L_b$
controls the motion of a relativistic particle
in the presence of Lorentz violation
involving the $b_\mu$ coefficients.
However,
identifying an application 
of the \rif\ $b$ structure $\fb$
appears challenging.

Some insight can be obtained by 
converting the variational problem associated with $\fb$
into a form with similarities to the \ran\ structure $\fa$.
This can be accomplished by introducing 
two additional coordinate variables,
a 2-form $\Si_{jk} = - \Si_{kj}$ and a scalar $\ka$,
and defining 
\beq
\fsb := 
\ssi + b^j \Si_{jk} \yy^k
+ \ka \ssi (\half \tr{\Si^2} +1) .
\label{fsb}
\eeq
Note that the factor of $\ssi$ in the last term
is included to maintain explicit homogeneity 
of degree one in $\yy^j$
but has no essential effect
on the argument to follow.
Note also that the conjugate velocities
for $\Si_{jk}$ and $\ka$ are absent from $\fsb$,
so an effective metric defined
in the enlarged space with coordinates $(x^j, \Si^{jk}, \ka)$
would have zero eigenvalues.

In the variational problem \rf{fsb},
the scalar $\ka$ plays the role of
a Lagrange multiplier,
enforcing the norm constraint 
$\Si_{jk}\Si^{jk} = 2$.
Variation with respect to the 2-form $\Si_{jk}$
imposes the condition
$b_j \yy_k - \yy_j b_k = 2\ka \ssi \Si_{jk}$.
These equations can be solved to yield
$\ka = \pm \be/{2\ssi}$
and $\Si_{jk} = \pm {(b_j \yy_k - b_k \yy_j)}/\be$,
which in turn can be used to show
that the canonical momentum $p_j$
associated with $x^j$ in $\fb$ 
coincides on shell with that in $\fsb$,
i.e.,
$p_j:= {\prt \fb}/{\prt \yy^j} = {\prt \fsb}/{\prt \yy^j}$.
It follows that $\fsb$ and $\fb$ have the same geodesics.
The two signs in $\fb$ 
correspond to $\ka>0$ and $\ka<0$ in $\fsb$.
Also, 
the condition $\ka = 0$ corresponds to $\be=0$
and hence for nonzero $\yy^j$ defines 
the set $S_1$ of singularities in $\tm S_0$.
A similar construction works for the
\prif\ structure $L_b$,
where the 2-form $\Si_{\mu\nu}$ takes the attributes
of the usual spin 2-tensor.

The expression \rf{fsb} reveals 
that for the $b$ structure 
the combination $b^k\Si_{kj}$ plays a role analogous
in certain respects to that of the \ran\ $a_j$ coefficient.
Since $\Si_{jk}$ is a dynamical variable,
this suggests $b$ space
can be viewed in terms of a \ran\ space
with a dynamical coefficient $a_j$.
Shen 
\cite{zs}
has shown that the usual \ran\ geodesics can be identified
with solutions to the Zermelo problem of navigation control 
in an external wind related to the coefficient $a_j$
(for a detailed exposition,
see the treatment by Bao and Robles \cite{dbcr}).
The dynamical coefficient $b^k \Si_{jk}$ 
therefore suggests a related interpretation for $b$ space
in which the effect of the external flow $b_j$ is adjustable,
in analogy to the change of effective wind direction 
arising from the combination of a boat's sail and keel.
A direct application to the Zermelo problem
falls short because in the \ran\ case
the external flow is related not only to the Zermelo wind 
but also to the \rie\ metric of the navigation space,
whereas the term $\ssi$ in the expression \rf{fsb}
is independent of $\Si_{jk}$.
However,
an interpretation of $\fb$ 
along these lines may be achievable 
for a system described by 
the more general theory of optimal control.

Another approach is to seek a physical system
in which the notion of distance is intrinsically quartic
rather than quadratic.
In the optical-metric interpretation,
for example,
the \ran\ structure $\fa$ generates geodesics 
matching the spatial trajectories
of null geodesics in a stationary spacetime,
which are determined by a quadratic spacetime interval
$ds^2 = 0$
\cite{ghww}.
In contrast,
geodesics of the $b$ structure $\fb$ 
match geodesics defined by a null quartic interval 
$ds^4 = 0$ in a certain class of spacetimes.
The quartic nature of $b$ space 
is directly reflected in its close ties
to the motion of a massive Dirac fermion,
which for nonzero $b_\mu$ generically 
has four distinct modes corresponding 
to the two spin degrees of freedom 
for particles and antiparticles,
whereas the spin-independent \ran\ case 
involves only two distinct modes.

\section*{Acknowledgments}

This work was supported in part
by the Department of Energy 
under grant DE-FG02-91ER40661
and by the Indiana University Center for Spacetime Symmetries.


\begin{thebibliography}{xx}

\bibitem{br}
B.\ Riemann,
{\it \"Uber die Hypothesen welche der Geometrie zu Grunde liegen},
in R.\ Baker, C.\ Christensen, and H.\ Orde,
{\it Bernhard Riemann, Collected Papers},
Kendrick Press, Heber City, Utah, 2004.

\bibitem{pf}
P.\ Finsler,
{\it \"Uber Kurven und Fl\"achen in allgemeinen R\"aumen},
University of G\"ottingen dissertation, 1918;
Verlag Birkh\"auser, Basel, Switzerland, 1951.

\bibitem{gr}
G.\ Randers, 
Phys.\ Rev.\ {\bf 59}, 195 (1941).

\bibitem{ksp}
V.A.\ Kosteleck\'y and S.\ Samuel,
Phys.\ Rev.\ D {\bf 39}, 683 (1989);
V.A.\ Kosteleck\'y and R.\ Potting,
Nucl.\ Phys.\ B {\bf 359}, 545 (1991).

\bibitem{kp}
V.A.\ Kosteleck\'y and R.\ Potting,
Phys.\ Rev.\ D {\bf 51}, 3923 (1995).

\bibitem{owg}
O.W.\ Greenberg,
Phys.\ Rev.\ Lett.\ {\bf 89}, 231602 (2002).

\bibitem{akgrav}
V.A.\ Kosteleck\'y,
Phys.\ Rev.\ D {\bf 69}, 105009 (2004).

\bibitem{cardinal}
V.A.\ Kosteleck\'y and R.\ Potting, 
Gen.\ Rel.\ Grav.\ {\bf 37}, 1675 (2005);
Phys.\ Rev.\ D {\bf 79}, 065018 (2009).

\bibitem{bcs}
D.\ Bao, S.-S.\ Chern, and Z.\ Shen,
{\it An Introduction to Riemann-Finsler Geometry},
Springer, New York, 2000.

\bibitem{cs}
S.-S.\ Chern and Z.\ Shen,
{\it Riemann-Finsler Geometry},
World Scientific, Singapore, 2005.

\bibitem{haz}
H.\ Akbar-Zadeh,
{\it Initiation to Global Finslerian Geometry},
North-Holland, Amsterdam, 2006.

\bibitem{bf}
A.\ Bejancu and H.R. Farran,
{\it Geometry of Pseudo-Finsler Submanifolds},
Kluwer Academic, Dordrecht, 2000.

\bibitem{ma}
R.\ Miron and M.\ Anastasiei,
{\it The Geometry of Lagrange Spaces: Theory and Applications},
Kluwer Academic, Dordrecht, 1994.

\bibitem{dcak}
D.\ Colladay and V.A.\ Kosteleck\'y,
Phys.\ Rev.\ D {\bf 55}, 6760 (1997);
Phys.\ Rev.\ D {\bf 58}, 116002 (1998).

\bibitem{gyb}
G.Yu.\ Bogoslovsky,
Phys.\ Lett.\ A {\bf 350}, 5 (2006);
SIGMA {\bf 1}, 017 (2005).

\bibitem{akrl}
V.A.\ Kosteleck\'y and R.\ Lehnert,
Phys.\ Rev.\ D {\bf 63}, 065008 (2001).

\bibitem{rl-fermion}
R.\ Lehnert,
J.\ Math.\ Phys.\ {\bf 45}, 2299 (2004).

\bibitem{badc}
B.\ Altschul and D.\ Colladay,
Phys.\ Rev.\  D {\bf 71}, 125015 (2005).

\bibitem{aknr}
V.A.\ Kosteleck\'y and N.\ Russell, 
Phys.\ Lett.\ B {\bf 693}, 443 (2010).

\bibitem{gls}
F.\ Girelli, S.\ Liberati, and L.\ Sindoni,
Phys.\ Rev.\ D {\bf 75}, 064015 (2007).

\bibitem{lld}
C.\ L\"ammerzahl, D.\ Lorek, and H.\ Dittus,
Gen.\ Rel.\ Grav.\ {\bf 41}, 1345 (2009).

\bibitem{km-nonmin}
V.A.\ Kosteleck\'y and M.\ Mewes,
Ap.\ J.\ Lett.\ {\bf 689}, L1 (2008);
Phys.\ Rev.\ D {\bf 80}, 015020 (2009).

\bibitem{gsa}
G.S.\ Asanov,
{\it Finsler Geometry, Relativity, and Gauge Theories},
Reidel, Dordtrecht, 1985.

\bibitem{sv}
S.I.\ Vacaru,
arXiv:0707.1526.

\bibitem{zcxl}
X.\ Li and Z.\ Chang,
arXiv:0711.1934;
Z.\ Chang and X.\ Li,
arXiv:0809.4762.

\bibitem{hg}
H.F.M.\ Goenner,
arXiv:0811.4529.

\bibitem{ls}
L.\ Sindoni,
Phys.\ Rev.\ D {\bf 77}, 124009 (2008).

\bibitem{kss}
A.P.\ Kouretsis, M.\ Stathakopoulos, and P.C.\ Stavrinos,
Phys.\ Rev.\ D {\bf 82}, 064035 (2010).

\bibitem{nm}
N.\ Mavromatos,
Phys.\ Rev.\ D {\bf 83}, 025018 (2010).

\bibitem{jsmv}
J.\ Sk\'akala and M.\ Visser,
Int.\ J.\ Mod.\ Phys.\ D {\bf 19}, 1119 (2010).

\bibitem{nv}
N.\ Voicu,
arXiv:1012.2100.

\bibitem{pw}
C.\ Pfeifer and M.N.R.\ Wohlfarth,
arXiv:1104.1079.

\bibitem{tables}
V.A.\ Kosteleck\'y and N.\ Russell,
Rev.\ Mod.\ Phys.\ {\bf 83}, 11 (2011).

\bibitem{jdak}
J.S.\ D\'\i az and V.A.\ Kosteleck\'y,
Phys.\ Lett.\ B, in press
[arXiv:1012.5985].

\bibitem{kvk}
V.A.\ Kosteleck\'y and R.\ Van Kooten,
Phys.\ Rev.\ D {\bf 82}, 101702(R) (2010).

\bibitem{fw}
L.L.\ Foldy and S.A.\ Wouthuysen,
Phys.\ Rev.\ {\bf 78}, 29 (1950).

\bibitem{qbakpn}
Q.G.\ Bailey and V.A.\ Kosteleck\'y,
Phys.\ Rev.\ D {\bf 74}, 045001 (2006).

\bibitem{lvmm}
R.\ Bluhm \etal, 
Phys.\ Rev.\ D {\bf 77}, 065020 (2008).

\bibitem{abk}
B.\ Altschul \etal,
Phys.\ Rev.\ D {\bf 81}, 065028 (2010).

\bibitem{mds}
M.D.\ Seifert,
Phys.\ Rev.\ Lett.\ {\bf 105}, 0201601 (2010);
Phys.\ Rev.\ D {\bf 82}, 125015 (2010).

\bibitem{akjt}
V.A.\ Kosteleck\'y and J.D.\ Tasson,
Phys.\ Rev.\ Lett.\ {\bf 102}, 010402 (2009);
Phys.\ Rev.\ D {\bf 83}, 016013 (2011).

\bibitem{hh}
H.\ Hironaka, 
Ann.\ Math. {\bf 79}, 109 (1964). 

\bibitem{aim}
P.L.\ Antonelli, R.S.\ Ingarden, and M.\ Matsumoto,
{\it The Theory of Sprays and Finsler Spaces
with Applications in Physics and Biology},
Kluwer Academic, Dordrecht, 1993.

\bibitem{cmm}
D.\ Colladay, P.\ McDonald, and D.\ Mullins,
J.\ Phys.\ A {\bf 43}, 275202 (2010).

\bibitem{diecke}
A.\ Deicke,
Arch.\ Math.\ {\bf 4}, 45 (1953).

\bibitem{mh}
M.\ Matsumoto,
Tensor, NS {\bf 24}, 29 (1972);
M.\ Matsumoto and S.\ H\=oj\=o,
Tensor, NS {\bf 32}, 225 (1978).

\bibitem{zs-sprays}
Z.\ Shen,
{\it Differential Geometry of Spray and Finsler Spaces},
Kluwer Academic, Dordrecht, 2001.

\bibitem{redef}
D.\ Colladay and P.\ McDonald,
J.\ Math.\ Phys.\ {\bf 43}, 3554 (2002);
M.S.\ Berger and V.A.\ Kosteleck\'y,
Phys.\ Rev.\ D {\bf 65}, 091701(R) (2002);
V.A.\ Kosteleck\'y and M.\ Mewes,
Phys.\ Rev.\ D {\bf 66}, 056005 (2002);
Q.G.\ Bailey and V.A.\ Kosteleck\'y,
Phys.\ Rev.\ D {\bf 70}, 076006 (2004);
B.\ Altschul,
J.\ Phys.\ A {\bf 39} 13757 (2006).

\bibitem{rl-redef}
R.\ Lehnert,
Phys.\ Rev.\ D {\bf 74}, 125001 (2006).

\bibitem{mm}
M.\ Matsumoto,
Kyoto Daigaku J.\ Math. {\bf 14}, 477 (1974).

\bibitem{hi}
M.\ Hashiguchi and Y.\ Ichijy\=o,
Rep.\ Fac.\ Sci.\ Kagoshima Univ.\ {\bf 8}, 39 (1975).

\bibitem{ssay}
C.\ Shibata, H.\ Shimada, M.\ Azuma, and H. Yasuda,
Tensor, NS {\bf 31}, 219 (1977).

\bibitem{sk}
S.\ Kikuchi,
Tensor, NS {\bf 33}, 242 (1979).

\bibitem{zs}
Z.\ Shen,
Canad.\ J.\ Math.\ {\bf 55}, 112 (2003).

\bibitem{brs}
D.\ Bao, C.\ Robles, and Z.\ Shen,
J.\ Diff.\ Geom.\ {\bf 66}, 377 (2004).

\bibitem{dbcr}
D.\ Bao and C.\ Robles,
in D.\ Bao, R.L.\ Bryant, S.-S.\ Chern, and Z.\ Shen, eds.,
{\it A Sampler of Riemann-Finsler Geometry},
Cambridge University Press, Cambridge, 2004.

\bibitem{ghww}
G.W.\ Gibbons, C.A.R.\ Herdeiro, C.M.\ Warnick, and M.C.\ Werner,
Phys.\ Rev.\ D {\bf 79}, 044022 (2009).

\bibitem{ts}
T.\ Sunada, 
Proc.\ KAIST Math.\ Workshop {\bf 8}, 93 (1993).

\end{thebibliography}
\end{document}